\documentclass[useAMS,usenatbib]{mn2e}

\usepackage{amsmath,texlogos}
\usepackage{natbib}
\usepackage{subfigure}
\usepackage{graphicx}
\usepackage{txfonts}
\usepackage[T1]{fontenc}
\usepackage{aecompl}
\usepackage{times}

\newcommand{\Msun}{\ensuremath{\,{\rm M}_\odot}}                  
\newcommand{\Rsun}{\ensuremath{\,{\rm R}_\odot}}                  
\newcommand{\Teff}{\ensuremath{T_{\rm eff}}}                      
\newcommand{\Mjup}{\ensuremath{\,{\rm M}_{\rm Jup}}}              
\newcommand{\Rjup}{\ensuremath{\,{\rm R}_{\rm Jup}}}              
\newcommand{\Teq}{\ensuremath{T_{\rm eq}^{\,\prime}}}             
\newcommand{\safronov}{\ensuremath{\Theta}}                       
\newcommand{\kms}{\,km\,s$^{-1}$}                                 
\newcommand{\ms}{\,m\,s$^{-1}$}                                   
\newcommand{\mss}{\,m\,s$^{-2}$}                                  
\newcommand{\as}{\ensuremath{^{\prime\prime}}}                    
\newcommand{\am}{\ensuremath{^\prime}}                            
\newcommand{\FeH}{\ensuremath{\left[\frac{\rm Fe}{\rm H}\right]}} 
\newcommand{\pjup}{\ensuremath{\,\rho_{\rm Jup}}}                 
\newcommand{\psun}{\ensuremath{\,\rho_\odot}}                     
\newcommand{\mc}[1]{\multicolumn{2}{c}{#1}}
\newcommand{\mcc}[1]{\multicolumn{3}{c}{#1}}

\begin{document}

\title[Photometry of exoplanets Qatar-1\,b and TrES-5\,b]{High-precision multi-band time-series photometry of exoplanets Qatar-1\,b and TrES-5\,b}
\author[D. Mislis]{D. Mislis$^{1}$\thanks{E-mail: dmislis@qf.org.qa}, L. Mancini$^{2}$, J. Tregloan-Reed$^{3}$, S. Ciceri$^{2}$, J. Southworth$^{4}$, G. D'Ago$^{5}$,
\newauthor
I. Bruni$^{6}$, \"O. Ba\c{s}t\"urk$^{7}$, K. A. Alsubai$^{1}$, E. Bachelet$^{1}$, D. M. Bramich$^{1}$, Th. Henning$^{2}$, 
\newauthor
T. C. Hinse$^{8}$, A. L. Iannella$^{5}$, N. Parley$^{1}$, T. Schroeder$^{2}$ \\
$^{1}$Qatar Environment and Energy Research Institute, Qatar Foundation, Tornado Tower, Floor 19, P.O. Box 5825, Doha, Qatar \\
$^{2}$Max Planck Institute for Astronomy, K\"onigstuhl 17, D-69117 Heidelberg, Germany\\
$^{3}$NASA Ames Research Center, Moffett Field, CA 94035, USA \\
$^{4}$Astrophysics Group, Keele University, Staffordshire ST5 5BG, UK \\
$^{5}$Department of Physics, University of Salerno, Via Giovanni Paolo II, 84084 Fisciano (SA), Italy \\
$^{6}$INAF – Osservatorio Astronomico di Bologna, Via Ranzani 1, I-40127 \\
$^{7}$Ankara University, Faculty of Science, Dept. of Astronomy and Space Sciences, Tando\u{g}an, TR-06100, Ankara, Turkey \\
$^{8}$Korea Astronomy and Space Science Institute, 776 Daedukdae-ro, Yuseong-gu, Daejeon 305-348, Republic of Korea \\
}

\maketitle

\date{Accepted ??. Received ??; in original form ??}

\pagerange{\pageref{firstpage}--\pageref{lastpage}} \pubyear{2014}

\label{firstpage}

\begin{abstract}
We present an analysis of the Qatar-1 and TrES-5 transiting exoplanetary systems, which contain Jupiter-like planets on short-period orbits around K-dwarf stars. Our data comprise 
a total of 20 transit light curves obtained using five medium-class telescopes, operated using the defocussing technique. The average precision we reach in all our data is 
$RMS_{Q} = 1.1$\,mmag for Qatar-1 ($V = 12.8$) and $RMS_{T} = 1.0$\,mmag for TrES-5 ($V = 13.7$). We use these data to refine the orbital ephemeris, photometric parameters, 
and measured physical properties of the two systems. One transit event for each object was observed simultaneously in three passbands ($gri$) using the BUSCA imager. The QES 
survey light curve of Qatar-1 has a clear sinusoidal variation on a period of $P_{\star} = 23.697 \pm 0.123$\,d, implying significant starspot activity. We searched for starspot 
crossing events in our light curves, but did not find clear evidence in any of the new datasets. The planet in the Qatar-1 system did not transit the active latitudes on the 
surfaces of its host star. Under the assumption that $P_{\star}$ corresponds to the rotation period of Qatar-1\,A, the rotational velocity of this star is very close to the 
$v \sin i_\star$ value found from observations of the Rossiter-McLaughlin effect. The low projected orbital obliquity found in this system thus implies a low absolute orbital 
obliquity, which is also a necessary condition for the transit chord of the planet to avoid active latitudes on the stellar surface.
\end{abstract}



\section{Introduction}

\begin{table*}
\caption{\label{tab:obs} Summary of observations of Qatar-1 and TrES-5.}
\centering
\begin{tabular}{lrrrrrrrrrr}
\hline
Telescope & Date & Start (UT) & End (UT) & Frames (No) & Exp (sec) & Filter & Airmass & Moon (\%) & Apertures (px) & RMS ($10^{-4}$) \\
\hline
Qatar-1: \\
CAHA 1.23m & 2011.08.25 & 23:55 & 03:36 & 158 & 60 & Cousins R & 1.20$\rightarrow$1.76 & 12.8 & 9,25,35 & 12.6 \\
CAHA 2.2m & 2011.08.25 & 23:46& 04:39 &115 &60 &Gunn g &1.19$\rightarrow$2.20 &12.8 &16.5,64.9,78.3& 10.5 \\
CAHA 2.2m &2011.08.25 &23:46& 04:39 &117 &60 &Gunn r& 1.19$\rightarrow$2.20 &12.8& 17,36.1,70 & 7.4 \\
CAHA 2.2m &2011.08.25 &23:46 &04:39 &118 &60 &Gunn i &1.19$\rightarrow$2.20 &12.8& 13.2,38.9,61.9& 9.6 \\
CAHA 1.23m & 2012.07.21 & 20:32 & 00:34 & 112& 120 & Cousins R & 1.33$\rightarrow$1.13 & 8.0 &7.9,31.9,45 & 7.1 \\
CAHA 1.23m & 2012.09.11 & 00:27 & 03:48 & 78 & 120 & Cousins R & 1.36$\rightarrow$2.20 & 27.1 & 19.9,38.9,53 & 7.5 \\
CAHA 1.23m & 2013.06.14 & 20:48 & 00:30 & 95 & 120 & Cousins R & 1.73$\rightarrow$1.19 & 19.3 &12.2,41,50.2 & 8.8 \\
CAHA 1.23m & 2013.07.28 & 20:17 & 01:05 & 156 & 120& Cousins R & 1.31$\rightarrow$1.16 & 59.1 &25.8,39,61 & 7.3  \\
CAHA 1.23m & 2014.04.19 & 01:40 & 04:38 & 109 & 160 & Cousins R & 1.61$\rightarrow$1.17 & 80.1 & 19,29,50 & 25.6 \\
CAHA 1.23m & 2014.06.04 & 20:42 & 02:57 & 138 & 150 & Cousins R & 1.97$\rightarrow$1.13 & 12.9 & 20,30,50 & 7.7 \\
TUG100 & 2014.06.04 & 21:39& 01:42& 98 &120 &Cousins R &1.69$\rightarrow$1.16& 12.9 &20,30,35 & 14.0 \\
CAHA 1.23m & 2014.09.07 & 23:57& 04:37& 135 &96-135 &Cousins R &1.25$\rightarrow$2.32& 98.2 &22,40,60 & 11.8 \\
\hline
TrES-5: \\
CAHA 2.2m &2011.08.26 &20:55 &00:55 &95& 60 &Gunn-g &1.02$\rightarrow$1.57 &7.0 &10.2,45.4,56.5& 12.3 \\
CAHA 2.2m &2011.08.26 &20:55 &00:55 &88& 60 &Gunn-r &1.02$\rightarrow$1.57 &7.0 &17.3,42.1, 55.6& 8.5 \\
CAHA 2.2m &2011.08.26 &20:55 &00:55 &91& 60 &Gunn-i &1.02$\rightarrow$1.57 &7.0 &10,28,39.1& 12.1 \\
CAHA 1.23m& 2012.09.10& 19:21& 23:36& 77& 170 &Cousins R& 1.11$\rightarrow$1.21 &29.0& 16,40.1,54.8& 8.5 \\
CAHA 1.23m& 2013.06.15& 00:42& 03:50& 65 &125 &Cousins R &1.14$\rightarrow$1.11 &44.9 &12,38,55.3 & 9.0 \\
CAHA 1.23m& 2013.07.30& 23:10& 04:03 &128 &120& Cousins I& 1.08$\rightarrow$1.47& 38.2 &13.7,27.8,46& 11.3 \\
Cassini 1.52m &2013.09.14 &21:19& 02:59 &139 &180& Gunn r& 1.08$\rightarrow$2.09& 78.2 &9.8,22.6,38.7& 13.0 \\
INT 2.5m & 2013.09.14 & 21:04 &22:22 &37&120 &Cousins I& 1.05$\rightarrow$1.17& 72.4& 18,28,50& 4.1 \\
\hline
\end{tabular}
\end{table*}

Ground-based photometric surveys have found a large number of transiting planets, possessing a huge diversity in their physical and orbital properties. The precise 
characterisation of these objects is a challenge as it requires high-quality data, both photometric and spectroscopic. The main limitation to our understanding of most 
transiting planets is due to the quality of the transit light curve, which is critical in determining the properties of both the planets and their host stars 
\citep{Southworth08,Southworth09}.\\
In this work we present follow-up photometry of two transiting planets orbiting cool stars -- Qatar-1\,b and TrES-5\,b -- aimed at improving measurements of their physical 
properties but also investigating the spot activity of their host stars. Our new data allow a significant improvement in our understanding of both systems, and, in the case of 
TrES-5, form the basis of the first study of the system since the discovery paper.\\
Qatar-1\,b was discovered by \citet{Alsubai+11}, and was the first planet found by the Qatar Exoplanet Survey (QES), an exoplanet transit survey focused on hot Jupiters and hot 
Neptunes via the transit method \citep{Alsubai+13}. The transiting planet TrES-5\,b was discovered shortly afterwards \citep{Mandushev+11} using observations by the TrES survey 
\citep{Alonso+04}. The Qatar-1 and TrES-5 systems are notably similar in terms of orbital period (1.4--1.5\,d), host star effective temperature (4800--5200\,K) and metallicity 
($\FeH = 0.20$), and the planetary radius ($\sim$1.2\Rjup) and equilibrium temperature (1400--1500\,K). Qatar-1 has subsequently been studied by \citet{Covino+13}, who found a 
sky-propjected orbital obliquity consistent with axial alignment, and by \citet{Essen+13}, who found indications of transit timing variations (TTVs) in this system. No studies 
of TrES-5 have been published since its discovery paper \citep{Mandushev+11}.\\
The possibility to observe occulted starspots during planetary-transit events opens new opportunities in the understanding of stars in general and planetary systems in 
particular. Those spots which are occulted by the planet manifest themselves as a small increase in flux during transit, which can be modelled to obtain the spot size, position 
and temperature \citep[e.g.][]{Mancini+14}. Multiple observations of the same spot during different transits can yield the orbital obliquity of the system 
\citep{Nutzman+11,Sanchis+11} to a significantly higher precision than achievable via the Rossiter-McLaughlin effect \citep{Tregloan+13}. The wavelength 
dependence of the amplitude of the unocculted starspots can mimic changes in the apparent radius of transiting planets as a function of wavelength \citep{Pont+13,Oshagh+14}.\\
In this work we present high-precision photometric observations of Qatar-1 and TrES-5, and use them to get more accurate measurements of the physical parameters of the systems. 
Some of our data were obtained in multiple passbands simultaneously, but we find no evidence for spot crossings in these data. We do, however, find strong evidence that the 
Qatar-1\,A is a spotted star from the long-term light curve of the system.


\section{Observations and data reduction}

\begin{figure*}
\centering
\includegraphics[width=15cm]{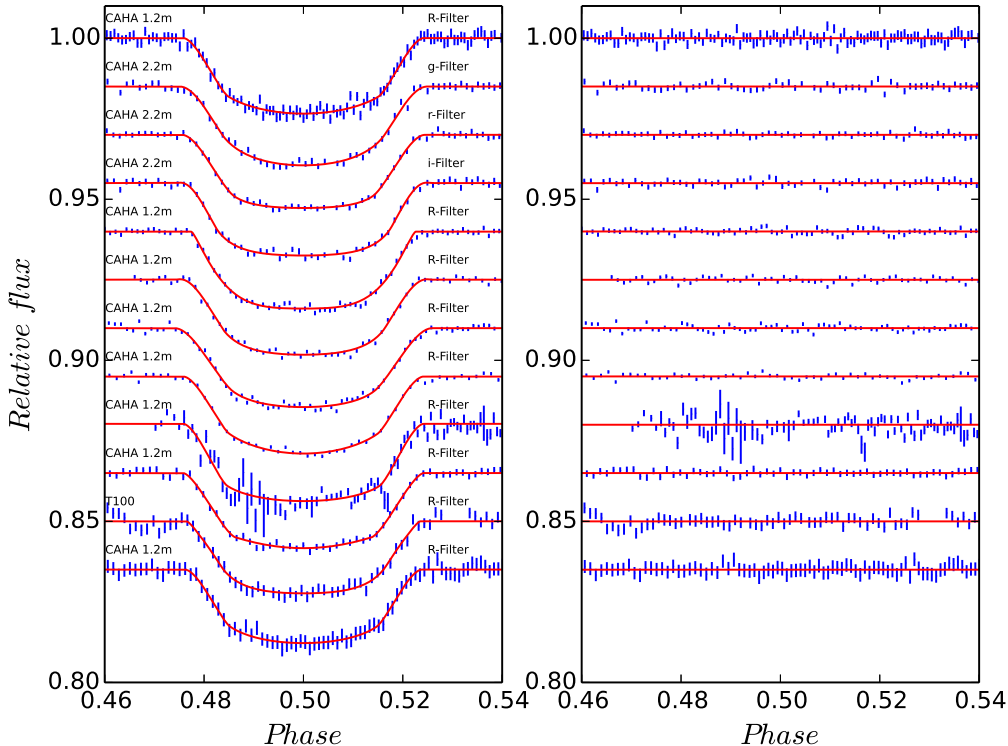}
\caption{Left: our light curves of Qatar-1 (blue points) with the {\sc jktebop} best fits (red lines) overplotted. The telescope and filter used for each dataset are labelled. 
Right: the residuals of each fit.}
\label{fig:q1:lc}
\end{figure*}
\noindent

Our observations were obtained using five medium-size telescopes equipped with imaging instruments, and operated out of focus (see \citealt{Southworth+09}). A summary of the observations is given in Table\,\ref{tab:obs}. The data were reduced using the {\sc defot} pipeline from \citet{Southworth+09,Southworth+14}. This pipeline was used to debias and flat-field the data, then perform aperture photometry on the target and all possible comparison stars. The radii of the software apertures (target, inner sky, outer sky) for each dataset were chosen to give the lowest scatter in the final light curve. The final light curve was constructed by calculating differential magnitudes versus a weighted set of comparison stars. The weights were optimised simultaneously with the coefficients of a low-order polynomial of magnitude versus time, in order to rectify the light curve to zero differential magnitude and minimise the scatter of the data obtained outside transit.

The data were reduced using the method and the {\sc defot} pipeline from \citet{Southworth+09,Southworth+14}.

A total of 11 transits were observed using the 1.23\,m telescope at Centro Astron\'omico Hispano-Alem\'an (CAHA). This uses a $2048\times 2048$ pixel CCD camera with a plate scale of 0.32\as\,px$^{-1}$ and has a $21.5\am\ \times 21.5\am$ field of view with the default $BVRI$ filters. Ten of the transits were obtained through a Cousins $R$ filter and the last through a Cousins $I$ filter. The first three transits were already presented in the study by \citet{Covino+13} but were re-reduced for the current work.

One transit each of Qatar-1 and TrES-5 was observed using the CAHA 2.2\,m telescope equipped with the BUSCA instrument. This obtains CCD images of a 5.8\am\ diameter field of view simultaneously in four optical passbands, split by dichroic elements. Each of the four CCDs has $4096 \times 4096$ pixels and is operated using 2$\times$2 binning. For both transits we obtained useful data in the Thuan-Gunn $g$, $r$ and $i$ passbands. The data taken through the Str\"omgren $u$ filter were discarded due to high scatter: both objects are comparatively faint ($V = 12.8$ for Qatar-1 and $V = 13.7$ for TrES-5) and cool so have very low flux levels in this passband.

One transit of Qatar-1 was monitored with the 1.0\,m telescope (T100) at T\"UB\.ITAK National Observatory (TUG) in Turkey, equipped with a $4096 \times 4096$ pixel CCD with a field of view of $21.5\am \times 21.5\am$. The transit was observed through a Cousins $R$ filter.

One transit of TrES-5 was obtained using the 2.5\,m Isaac Newton Telescope (INT) at La Palma, Spain, and the Wide Field Camera (WFC). This is a mosaic of four CCDs of which we used only CCD4, to avoid possible systematic errors and calibration issues resulting from the use of multiple CCDs in the mosaic. This CCD has $2048\times 4096$ pixels, giving a field of view of $11.3\am \times 22.5\am$, and a Cousins $I$ filter was selected.

Finally, a transit of TrES-5 was observed using the Cassini 1.52\,m telescope at Bologna Astronomical Observatory, Loiano, Italy. The BFOSC instrument was used in imaging mode, with a Thuan-Gunn $r$ filter. The $1024\times 1024$ pixels CCD provided a field of view of $13\am \times 12.6\am$ at 0.58\as\ per pixel.


\section{Transit analysis}

Each transit light curve was modelled with the {\sc jktebop} code to extract measurements of its photometric parameters. The object size parameters in {\sc jktebop} are the fractional radii of the star and the planet ($r_{\rm A}$ and $r_{\rm b}$), which are the ratios between the true radii and the semimajor axis ($r_{\rm A,b} = \frac{R_{\rm A,b}}{a}$). The fitted parameters were the sum of the fractional radii ($r_{\rm A} + r_{\rm b}$), the ratio of the radii ($k = \frac{r_{\rm b}}{r_{\rm A}} = \frac{R_{\rm b}}{R_{\rm A}}$), the orbital inclination ($i$), and a reference time of mid-transit. We assumed an orbital eccentricity of zero for both objects based on previous studies \citep{Covino+13,Mandushev+11}. Limb darkening was applied using the quadratic law, with coefficients taken from \citet{Claret04}. We used Monte Carlo simulations to perform the error analysis for each transit fit. The errors were propagated following \citet{Alonso+08} and \citet{Mislis+10}.

\subsection{Qatar-1}

\begin{table*}
\centering
\caption{\label{tab:q1:lc} Fitted parameter values for each light curve of Qatar-1.}
\begin{tabular}{lcccccc}
\hline
Date & $r_{\rm A}+r_{\rm b}$ & $k$ & $r_{\rm A}$ & $r_{\rm b}$ & Inclination ($^\circ$) & $T_0$ (BJD/TDB)  \\
\hline
2011.08.25 & $0.178 \pm 0.011$ & $0.1455 \pm 0.0039$ & $0.155 \pm 0.009$ & $0.0226 \pm 0.0018$ & $84.64 \pm 0.84$ & $55799.5759 \pm 0.0002$ \\ 
2011.08.25 & $0.179 \pm 0.009$ & $0.1469 \pm 0.0039$ & $0.156 \pm 0.007$ & $0.0229 \pm 0.0016$ & $84.62 \pm 0.66$ & $55799.5755 \pm 0.0001$ \\ 
2011.08.25 & $0.184 \pm 0.009$ & $0.1467 \pm 0.0041$ & $0.160 \pm 0.007$ & $0.0235 \pm 0.0016$ & $84.02 \pm 0.73$ & $55799.5758 \pm 0.0002$ \\ 
2011.08.25 & $0.172 \pm 0.011$ & $0.1428 \pm 0.0031$ & $0.150 \pm 0.009$ & $0.0215 \pm 0.0018$ & $85.18 \pm 0.89$ & $55799.5756 \pm 0.0001$ \\ 
2012.07.21 & $0.185 \pm 0.004$ & $0.1500 \pm 0.0022$ & $0.161 \pm 0.004$ & $0.0241 \pm 0.0008$ & $83.95 \pm 0.32$ & $56130.4430 \pm 0.0001$ \\
2012.09.11 & $0.178 \pm 0.007$ & $0.1461 \pm 0.0028$ & $0.155 \pm 0.006$ & $0.0226 \pm 0.0010$ & $84.50 \pm 0.52$ & $56181.5668 \pm 0.0001$ \\ 
2013.06.14 & $0.197 \pm 0.006$ & $0.1524 \pm 0.0024$ & $0.171 \pm 0.004$ & $0.0260 \pm 0.0010$ & $83.33 \pm 0.34$ & $56458.4665 \pm 0.0002$ \\ 
2013.07.28 & $0.183 \pm 0.006$ & $0.1480 \pm 0.0020$ & $0.159 \pm 0.004$ & $0.0235 \pm 0.0010$ & $84.10 \pm 0.42$ & $56502.4905 \pm 0.0002$ \\  
2014.04.19 & $0.176 \pm 0.018$ & $0.1471 \pm 0.0083$ & $0.153 \pm 0.015$ & $0.0226 \pm 0.0030$ & $84.91 \pm 1.50$ & $56766.6120 \pm 0.0004$ \\  
2014.06.04 & $0.181 \pm 0.009$ & $0.1458 \pm 0.0028$ & $0.158 \pm 0.007$ & $0.0231 \pm 0.0014$ & $84.41 \pm 0.66$ & $56813.4731 \pm 0.0002$ \\ 
2014.06.04 & $0.182 \pm 0.020$ & $0.1446 \pm 0.0076$ & $0.159 \pm 0.016$ & $0.0230 \pm 0.0032$ & $84.10 \pm 1.50$ & $56813.4720 \pm 0.0003$ \\ 
2014.09.07 & $0.173 \pm 0.011$ & $0.1427 \pm 0.0035$ & $0.151 \pm 0.009$ & $0.0216 \pm 0.0018$ & $83.86 \pm 0.80$ & $56908.6149 \pm 0.0002$ \\ 
\hline

Weighted mean & $0.184 \pm 0.002$ & $0.1475 \pm 0.0009$ & $0.160 \pm 0.002$ & $0.0236 \pm 0.0004$ & $84.03 \pm 0.16$ \\

\hline
\end{tabular}
\end{table*}

\begin{figure}
\includegraphics[width=\columnwidth]{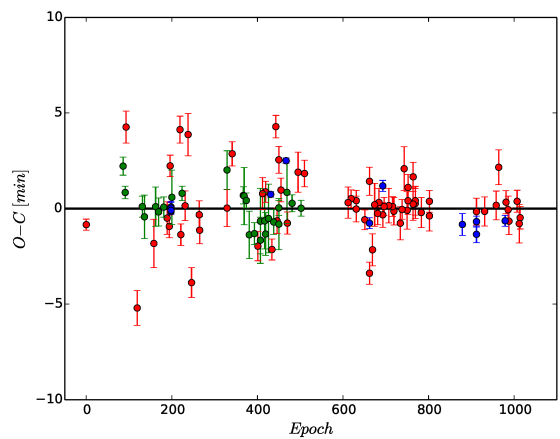}
\caption{$O-C$ diagram for the transit times of Qatar-1. Our own data are shown using blue dots, 26 data points from \citet{Essen+13} (green dots), and the 69 additional light curves 
from ETD database (red dots).}
\label{fig:q1:ttv}
\end{figure}

For Qatar-1, we collected 12 light curves in total (see Fig.\,\ref{fig:q1:lc}). We fit each of the datasets individually, obtaining the parameter values given in Table\,\ref{tab:q1:lc}. The parameter values in Table\,\ref{tab:q1:lc} were combined into weighted means for the determination of the physical properties of the system (see below). We then fitted the $T_0$ values with a straight line versus cycle number to determine the orbital ephemeris. The uncertainties were obtained using 1000 Monte Carlo simulations. The resulting ephemeris is:
\begin{equation}
T_0 = 2455799.57954 (4) + 1.42002586 (275) \cdot E \label{eq01}
\end{equation}
where $T_0$ is the transit mid-time, $E$ is the cycle number and the bracketed quantities give the uncertainty in the final digit of the preceding number. All times in our analysis were converted to Barycentric Julian Day (BJD/TDB).

We supplemented our $T_0$ values with data from the literature and searched for TTVs. We included timings from the ETD amateur database\footnote{\tt http://var2.astro.cz/ETD/} with quality higher than 3. We fit a linear function to $T_0$ and then removed the linear trend. Fig.\,\ref{fig:q1:ttv} shows the results ($O-C$ diagram) overplotted with the best linear fit. The $\chi_{red}^{2}$ value is 31.4, which is very high. This implies that the $O-C$ data cannot be explained by a simple linear fit, but still the amplitude of our O-C residuals are smaller ($RMS_{O-C} = 1.50$ minutes) than \citet{Essen+13} ($RMS_{O-C} = 1.67$ minutes), \citet{Covino+13} ($RMS_{O-C} = 2.45$ minutes) or ETD ($RMS_{O-C} = 3.85$ minutes). \citet{Essen+13} found evidence for TTVs in Qatar-1 but we need further and more precise data in order to analyse this scenario in detail.

\subsection{Multi-band photometry of Qatar-1}

\begin{figure}
\includegraphics[width=8cm]{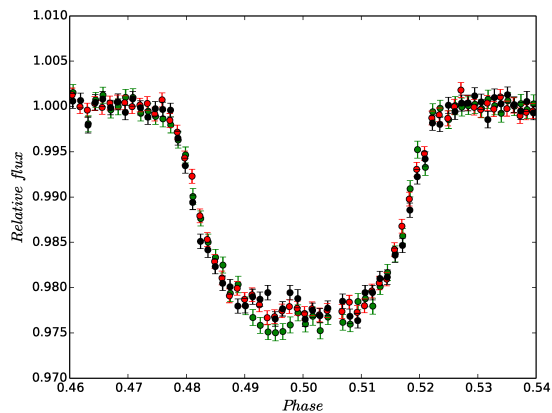}
\caption{The BUSCA light curves of Qatar-1. Green, red and black points show the $g$, $r$ and $i$ data, respectively.}
\label{fig:q1:busca}
\end{figure}

\begin{figure}
\includegraphics[width=\columnwidth]{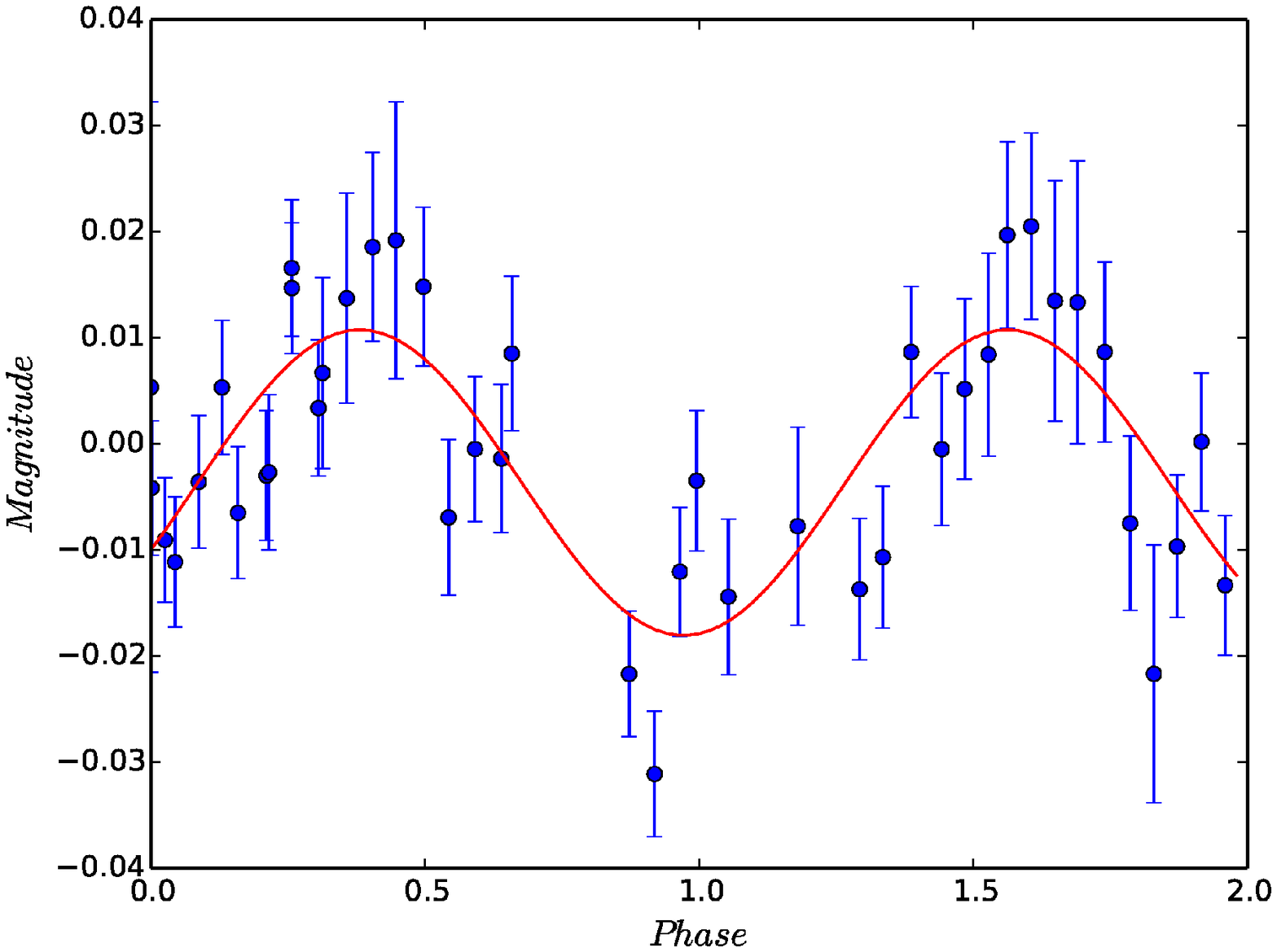}
\caption{Long-term light curve of Qatar-1 from QES, phase-folded
on a period of 23.697\,d. The best fit is shown by a red line.}
\label{fig:q1:qes}
\end{figure}

The BUSCA light curves were obtained simultaneously in three filters using the same telescope and instrument, so are useful for investigating the possible presence of starspots. Fig.\,\ref{fig:q1:busca} shows the three light curves overplotted. Whilst there are suggestions of starspots in the $g$ and $r$ data, these are close to the level of the noise so their existence is not proven. This transit was also monitored using the CAHA 1.23\,m telescope (first dataset in Fig.\,\ref{fig:q1:lc}), and these data do not confirm the presence of any starspots.

Whilst we have no clear detection of starspots via occultation during transit, spots are a common phenomenon on the surfaces of K-type dwarfs. They can cause brightness modulation at the rotational period (and/or its submultiples) of the star. We used the discovery light curves from QES \citep{Alsubai+11}, which span 380 days, to search for stellar variability. A Lomb-Scargle periodogram of the data shows a clear detection of sinusoidal modulation at a period of $P_{\star} = 23.697 \pm 0.123$\,d (Fig.\,\ref{fig:q1:qes}), which we take to be the rotational period of the star. The implied stellar rotation velocity of $v = 1.76$\kms\ is fully consistent with the value of $v \sin i_{\star} = 1.7 \pm 0.3$\kms\ found by \citet{Covino+13} from the Rossiter-McLaughlin effect. This in turn indicates that the inclination of the stellar rotation axis, $i_{\star}$, is close to 90$^\circ$, so the {\em true} orbital obliquity of the system is close to the measured value of the {\em projected} orbital obliquity found by \citet{Covino+13}.


\subsection{TrES-5b}

\begin{figure*}
\includegraphics[width=\textwidth]{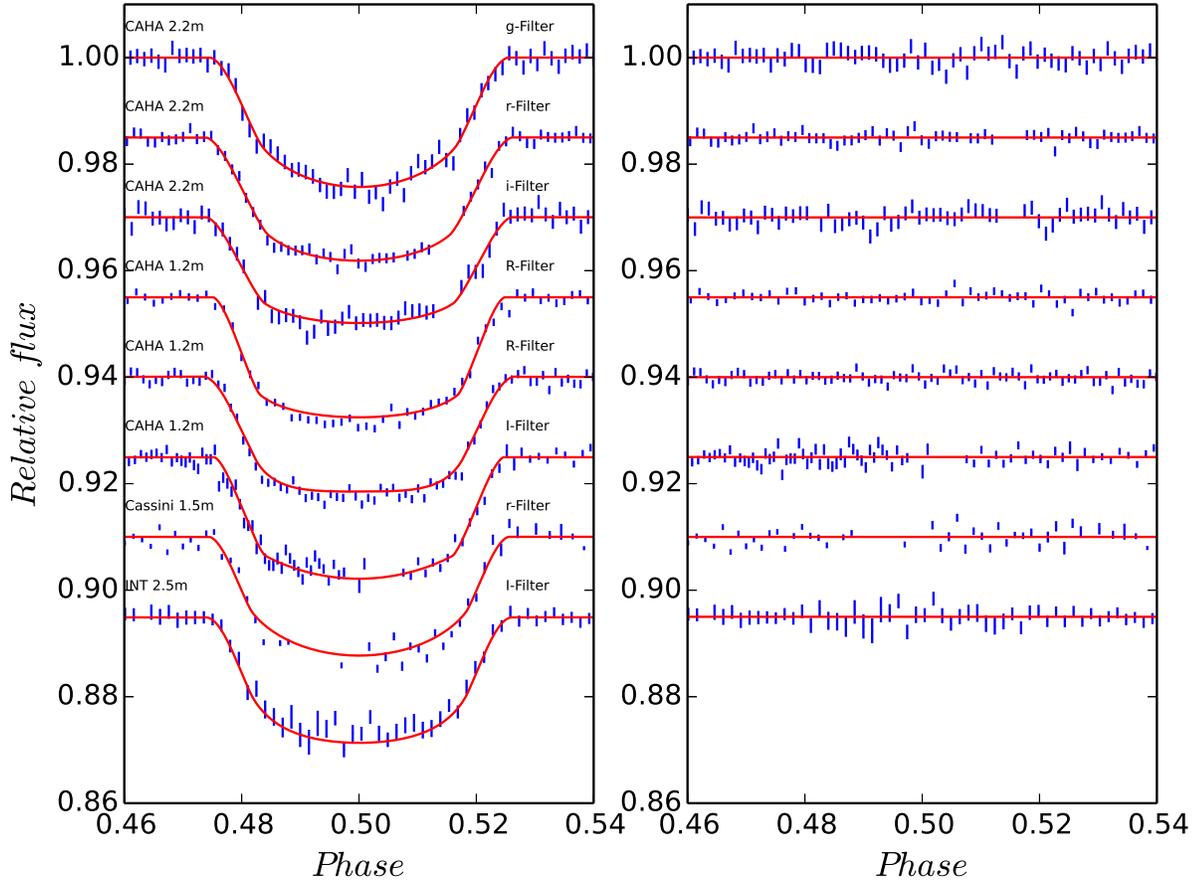}
\caption{Left: our light curves of TrES-5 (blue points) with the {\sc jktebop} best fits (red lines)
overplotted. The telescope and filter used for each dataset are labelled. Right: the residuals of each fit.}
\label{fig:t5:lc}
\end{figure*}

\begin{table*}
\centering
\caption{\label{tab:t5:lc} Fitted parameter values for each light curve of TrES-5.}
\begin{tabular}{lcccccc}
\hline
Date & $r_{\rm A}+r_{\rm b}$ & $k$ & $r_{\rm A}$ & $r_{\rm b}$ & Inclination ($^\circ$) & $T_0$ (BJD/TDB)  \\
\hline
2011.08.26 & $0.180 \pm 0.016$ & $0.144 \pm 0.007$ & $0.157 \pm 0.013$ & $0.0227 \pm 0.0028$ & $85.04 \pm 1.42$ & $55800.4753 \pm 0.0002$ \\
2011.08.26 & $0.198 \pm 0.006$ & $0.146 \pm 0.002$ & $0.172 \pm 0.005$ & $0.0252 \pm 0.0011$ & $83.58 \pm 0.42$ & $55800.4744 \pm 0.0002$ \\
2011.08.26 & $0.194 \pm 0.014$ & $0.135 \pm 0.005$ & $0.171 \pm 0.011$ & $0.0232 \pm 0.0022$ & $83.79 \pm 0.90$ & $55800.4751 \pm 0.0003$ \\
2012.09.10 & $0.184 \pm 0.014$ & $0.139 \pm 0.005$ & $0.161 \pm 0.011$ & $0.0225 \pm 0.0022$ & $84.87 \pm 0.98$ & $56181.4120 \pm 0.0002$ \\
2013.06.15 & $0.188 \pm 0.008$ & $0.145 \pm 0.002$ & $0.164 \pm 0.007$ & $0.0238 \pm 0.0015$ & $84.55 \pm 0.58$ & $56458.5923 \pm 0.0001$ \\
2013.07.30 & $0.170 \pm 0.010$ & $0.141 \pm 0.005$ & $0.149 \pm 0.008$ & $0.0210 \pm 0.0015$ & $85.10 \pm 0.84$ & $56504.5419 \pm 0.0001$ \\
2013.09.14 & $0.184 \pm 0.014$ & $0.145 \pm 0.005$ & $0.161 \pm 0.011$ & $0.0232 \pm 0.0002$ & $85.48 \pm 0.91$ & $56550.4919 \pm 0.0001$ \\
2013.09.14 & $0.180 \pm 0.014$ & $0.139 \pm 0.005$ & $0.158 \pm 0.011$ & $0.0220 \pm 0.0022$ & $84.26 \pm 1.03$ & $56550.4915 \pm 0.0001$ \\

\hline
Weighted mean & $0.188 \pm 0.004$ & $0.143 \pm 0.0012$ & $0.164 \pm 0.003$ & $0.0232 \pm 0.0002$ & $84.27 \pm 0.26$ \\

\hline

\hline
\end{tabular}
\end{table*}

\begin{figure}
\includegraphics[width=\columnwidth]{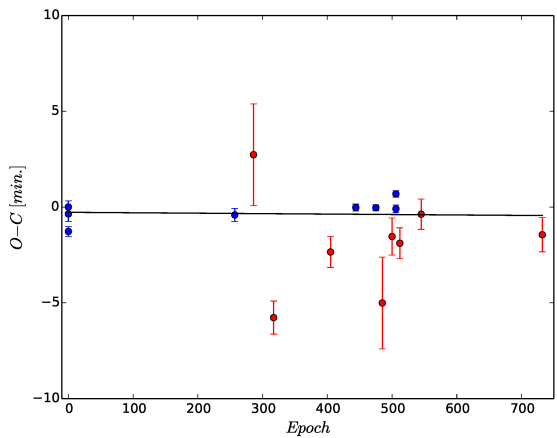}
\caption{$O-C$ diagram for the transit times of TrES-5. Our own data are shown using
blue dots and the eight additional light curves from ETD database are shown with red dots.}
\label{fig:t5:ttv}
\end{figure}

\begin{figure}
\includegraphics[width=\columnwidth]{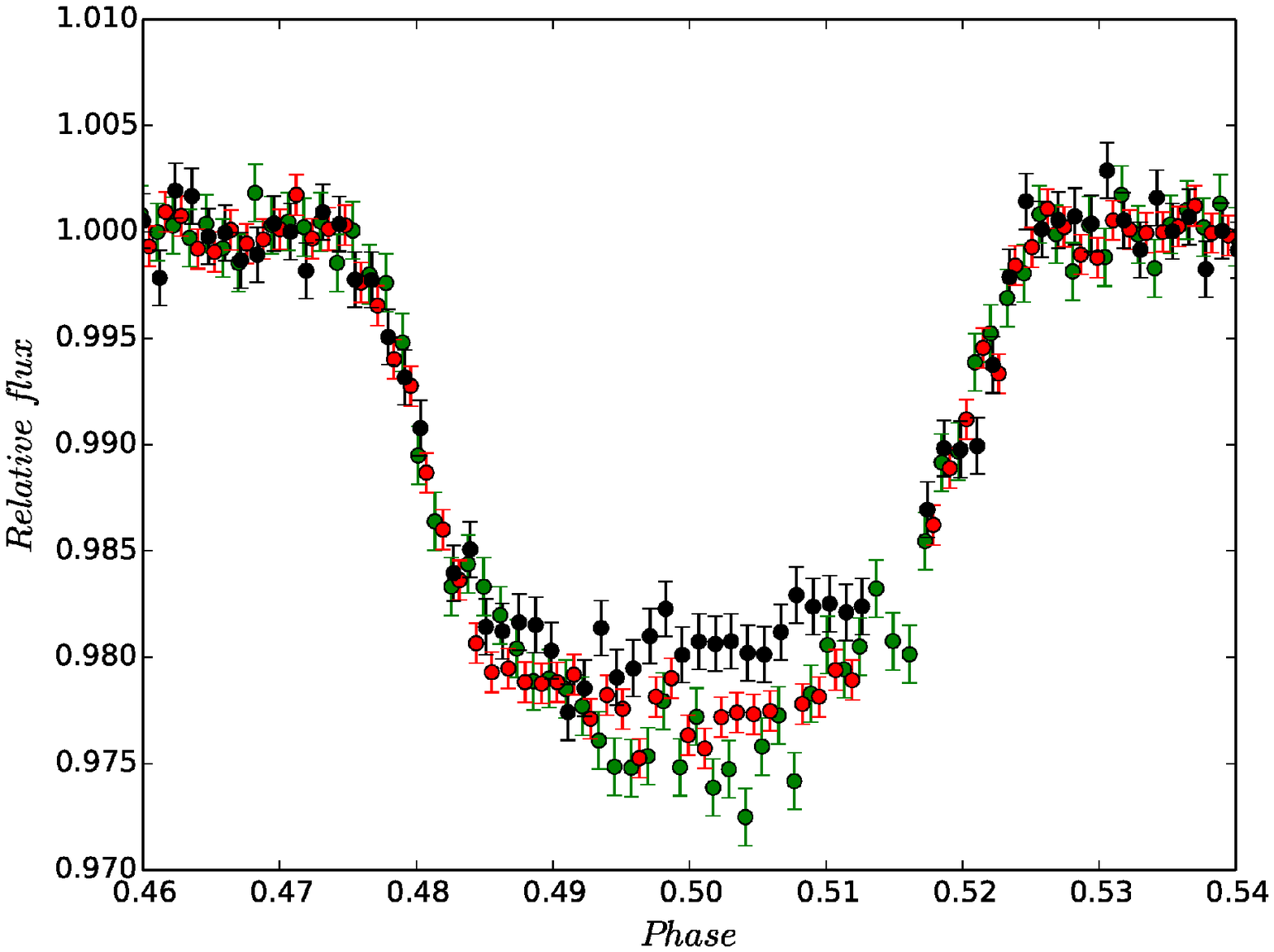}
\caption{The BUSCA light curves of TrES-5. Green, red and
black points show the $g$, $r$ and $i$ data, respectively.}
\label{fig:t5:busca}
\end{figure}

The analysis of our eight light curves of TrES-5 followed the same steps as for Qatar-1 above. The best-fitting photometric parameters are given in Table\,\ref{tab:t5:lc} and the best fits are plotted in Fig.\,\ref{fig:t5:lc}. The parameter values in Table\,\ref{tab:t5:lc} were combined into weighted means for the determination of the physical properties of the system (see below). The resulting orbital ephemeris is:
\begin{equation}
T_0 = 2456458.59219 (9) + 1.48224686 (614) \cdot E \label{eq02}
\end{equation}
As with Qatar-1, we added $T_0$ measurements from the ETD database (again using only those with qualities higher than 3) and formed the $O-C$ diagram (Fig.\,\ref{fig:t5:ttv}). The best-fitting ephemeris has $\chi_{red}^{2} = 7.15$, which a factor of ten lower than that for Qatar-1. This $\chi_{red}^{2}$ indicates either that the linear ephemeris is a poor representation of the data or that the errorbars of many of the $T_0$ values are underestimated. As with Qatar-1, further data are needed to investigate this situation and to provide clear evidence (or otherwise) of the presence of TTVs in this system.

The multi-band data from BUSCA are shown in Fig.\,\ref{fig:t5:busca}, and contain no clear evidence of starspot occultations. The data during totality (between 2nd and 3rd contact) are rather noisy so are not good indicators of the presence of starspots. This is at least partly due to the relative faintness of TrES-5 ($V = 13.7$), at least for a 2-m class telescope equipped with an instrument containing many optical elements.


\section{Physical properties}

\begin{table}
\centering
\caption{\label{tab:spec} Spectroscopic properties of the host stars
in the Qatar-1 and TrES-5 systems adopted from the literature.
\newline {\bf References:} (1) \citet{Covino+13}, (2) \citet{Mandushev+11}.}
\begin{tabular}{l r@{\,$\pm$\,}l c r@{\,$\pm$\,}l c}
\hline
Source            & \mc{Qatar-1} & Ref & \mc{TrES-5} & Ref \\
\hline
\Teff\ (K)        & 4910  & 100  & 1   & 5171  & 36   & 2  \\
\FeH\ (dex)       & 0.20  & 0.10 & 1   & 0.20  & 0.10 & 2  \\
$K_{\rm A}$ (\ms) & 265.7 & 3.5  & 1   & 339.8 & 10.4 & 2  \\
\hline
\end{tabular}
\end{table}

\begin{table*}
\centering
\caption{\label{tab:model} Derived physical properties of the two systems. Where two sets of errorbars
are given, the first is the statistical uncertainty and the second is the systematic uncertainty.}
\begin{tabular}{l l l r@{\,$\pm$\,}c@{\,$\pm$\,}l r@{\,$\pm$\,}c@{\,$\pm$\,}l}
\hline
Quantity                & Symbol           & Unit    & \mcc{Qatar-1}                   & \mcc{TrES-5}                    \\
\hline
Stellar mass            & $M_{\rm A}$      & \Msun  & 0.818    & 0.047    & 0.050      & 0.901    & 0.029    & 0.008     \\
Stellar radius          & $R_{\rm A}$      & \Rsun  & 0.796    & 0.016    & 0.017      & 0.868    & 0.013    & 0.002     \\
Stellar surface gravity & $\log g_{\rm A}$ & c.g.s. & 4.549    & 0.011    & 0.009      & 4.517    & 0.012    & 0.001     \\
Stellar density         & $\rho_{\rm A}$   & \psun  & \mcc{$1.621 \pm 0.046$}          & \mcc{$1.381 \pm 0.051$}           \\[2pt]
Planet mass             & $M_{\rm b}$      & \Mjup  & 1.293    & 0.052    & 0.054      & 1.790    & 0.067    & 0.010     \\
Planet radius           & $R_{\rm b}$      & \Rjup  & 1.142    & 0.026    & 0.024      & 1.194    & 0.015    & 0.003     \\
Planet surface gravity  & $g_{\rm b}$      & \mss   & \mcc{$24.56 \pm  0.70$}            & \mcc{$31.1 \pm  1.0$}           \\
Planet density          & $\rho_{\rm b}$   & \pjup  & 0.811    & 0.036    & 0.017      & 0.983    & 0.039    & 0.003     \\[2pt]
Equilibrium temperature & \Teq\            & K      & \mcc{$1388 \pm   29$}            & \mcc{$1480 \pm   13$}           \\
Safronov number         & \safronov\       &        & 0.0640   & 0.0017   & 0.0014     & 0.0817   & 0.0028   & 0.002    \\
Orbital semimajor axis  & $a$              & au     & 0.02313  & 0.00044  & 0.00048    & 0.02459  & 0.00026  & 0.0007   \\
Age (gyrochronology)    & $\tau_{\rm g}$   & Gyr    & \mcc{$1.19 \pm 0.47$}                                           \\[2pt]
\hline
\end{tabular}
\end{table*}

We determined the physical properties of the two systems from the light curve fits (the weighted means in Tables \ref{tab:q1:lc} and \ref{tab:t5:lc}), from the spectroscopic measurements of host star's atmospheric properties, and from the tabulated predictions of five different sets of theoretical stellar evolutionary models. The values of $r_{\rm A}$, $r_{\rm b}$ and $i$ in Tables \ref{tab:q1:lc} and \ref{tab:t5:lc} were combined according to their weighted mean, inflating the resulting errorbars to enforce $\chi^2_\nu = 1.0$ for each quantity. The spectroscopic measurements of the stellar effective temperature (\Teff), metallicity $\left( \FeH \right)$ and orbital velocity amplitude ($K_{\rm A}$) were taken from published studies and are summarised in Table\,\ref{tab:spec}. Tabulated predictions were obtained from the \citet{Claret04a}, Y$^2$ \citep{Demarque+04}, Teramo \citep{Pietrinferni+04}, VRSS \citep{VandenBerg+06} and DSEP \citep{Dotter+08} stellar models.

For each target we began by estimating a value for the velocity amplitude of the {\em planet}, $K_{\rm b}$, allowing us to calculate a set of physical properties for the system using standard formulae. The value of $K_{\rm b}$ was then iteratively refined to maximise the agreement between the observed and predicted \Teff, and the measured $r_{\rm A}$ and predicted $\frac{R_{\rm A}}{a}$. Finally, the full procedure was undertaken using each of the five sets of stellar model predictions. For both objects we assumed a circular orbit, based on the conclusions of \citet{Covino+13} and \citet{Mandushev+11}. We used the set of physical constants given by \citet{Southworth11}.

The uncertainties on the input parameters were propagated through the analysis using a perturbation approach, and added in quadrature to give the final random error. A systematic errorbar was also estimated based on the interagreement between the results obtained using each of the five different model sets. Table\,\ref{tab:model} gives our final physical properties, random errorbars for all quantities, and systematic errorbars for those results which depend on stellar theory.

Also, only for Qatar-1A, we were able to calculate the stellar rotation period. Thus, we can use gyrochronology model to estimate stellar the age of Qatar-1b host star. Using the model from \citet{Barnes+07} and stellar rotation period 23.697 days and B-V=1.06, we estimate the age of Qatar-1A $\tau _{g} =1.1865 \pm 0.47$ Gyr.

Our final results for Qatar-1 are in good agreement with published studies \citep{Alsubai+11,Covino+13}, and yield a significant improvement in precision. In the case of TrES-5 we agree with the findings of \citet{Mandushev+11} but do not obtain significantly smaller errorbars. This is because we account for systematic errors whereas \citet{Mandushev+11} do not, and also because the errors estimated by \citet{Mandushev+11} appear to be too small (for example they claim that the orbital inclination is $i = 84.529 \pm 0.005$ degrees, a level of precision not normally achieved even with {\it Kepler} or {\it Hubble Space Telescope} light curves). Our results are therefore to be preferred to those of \citet{Mandushev+11} because they are based on a larger and more precise dataset, and because our errorbars have been more robustly calculated.


\section{Results and Conclusions}

We present extensive optical photometry of transit events in two extrasolar planetary systems with K-dwarf host stars. Our data comprise 12 light curves of Qatar-1 and eight light curves of TrES-5. These data include simultaneous observations in three passbands of one transit for each object. We use thse data to search for starspot crossing events during transit, with a negative result. We do, however, measure the rotational period of $P_{\star} = 23.697 \pm 0.123$\,d for Qatar-1\,A from the survey photometry in its discovery paper, showing that this does display spot activity. The corresponding rotational velocity is close to the $v \sin i_\star$ value measured from an observation of the Rossiter-McLaughlin effect in this system, so its low projected orbital obliquity also implies a low true orbital obliquity. The lack of observed spot crossings may be due to the planets crossing latitudes of the stars which show low spot activity, i.e.\ the planetary chords miss the active latitudes of the stellar surfaces.

We use our data to measure the photometric parameters of both systems. When combined with published spectroscopic quantities, these yield precise measurements of the full physical properties of the systems. Qatar-1 and TrES-5 have notable similarities in their respective stellar properties, and planetary equilibrium temperature, radius and density. Our results also yield refined measurements of the orbital ephemerides of the systems.


\section*{Acknowledgments}

This publication is supported by NPRP grant $\sharp$ X-019-1-006 from the Qatar National Research Fund (a member of Qatar Foundation). The statements made herein are solely the responsibility of the authors. \\
Based on observations obtained with the 1.52-m Cassini telescope at the OAB Observatory in Loiano (Italy), and with the 1.23-m and 2.2m telescopes at the Centro Astron\`{o}mico 
Hispano Alem\`{a}n (CAHA) at Calar Alto (Spain), jointly operated by the Max-Planck Institut for Astronomy and the Instituto de Astrof\`{\i}sica de Andaluc\`{\i}a (CSIC).\\
We thank to  T\"{U}B\.{I}TAK for the partial support in using T100 telescope with project number 12CT100-378. OB acknowledges the support by the research fund of Ankara 
University (BAP) through the project 13B4240006.


\bsp

\label{lastpage}


\begin{thebibliography}{99}
\bibitem[\protect\citeauthoryear{Alonso et al.}{2004}]{Alonso+04} Alonso R., et al., 2004, ApJ, 613, L153                                                   %
\bibitem[\protect\citeauthoryear{Alonso et al.}{2008}]{Alonso+08} Alonso R., et al.,2008, A\&A, 487, L5                                                     
\bibitem[\protect\citeauthoryear{Alsubai et al.}{2011}]{Alsubai+11} Alsubai K. A., et al., 2011, MNRAS, 417, 709                                            
\bibitem[\protect\citeauthoryear{Alsubai et al.}{2013}]{Alsubai+13} Alsubai K. A., et al., 2013, AcA, 63, 465                                               
\bibitem[\protect\citeauthoryear{Barnes}{2007}]{Barnes+07} Barnes S., 2007, ApJ, 669, 1167B
\bibitem[\protect\citeauthoryear{Claret}{2004a}]{Claret04a} Claret A., 2004, A\&A, 424, 919                                                                 %
\bibitem[\protect\citeauthoryear{Claret}{2004b}]{Claret04} Claret A., 2004, A\&A, 428, 1001                                                                 
\bibitem[\protect\citeauthoryear{Covino et al.}{2013}]{Covino+13} Covino E., et al., 2013, A\&A, 554, A28                                                   
\bibitem[\protect\citeauthoryear{Demarque et al.}{2004}]{Demarque+04} Demarque P., et al., 2004, ApJS, 155, 667                                             
\bibitem[\protect\citeauthoryear{Dotter et al.}{2008}]{Dotter+08} Dotter A., et al., 2008, ApJS, 178, 89                                                    
\bibitem[\protect\citeauthoryear{Mancini et al.}{2014}]{Mancini+14} Mancini L., et al. 2014, MNRAS, 443, 2391                                               %
\bibitem[\protect\citeauthoryear{Mandushev et al.}{2011}]{Mandushev+11} Mandushev G., et al., 2011, ApJ, 741, 114                                           
\bibitem[\protect\citeauthoryear{Mislis et al.}{2010}]{Mislis+10} Mislis D., et al., 2010, A\&A, 510, A107                                                  
\bibitem[\protect\citeauthoryear{Nutzman et al.}{2011}]{Nutzman+11} Nutzman P.\ A., Fabrycky D.\ C., Fortney J.\ J., 2011, ApJ, 740, L10                    %
\bibitem[\protect\citeauthoryear{Oshagh et al.}{2014}]{Oshagh+14} Oshagh M., et al., 2014, A\&A, 568, A99                                                   %
\bibitem[\protect\citeauthoryear{Pietrinferni et al.}{2004}]{Pietrinferni+04} Pietrinferni A., et al., 2004, ApJ, 612, 168                                  
\bibitem[\protect\citeauthoryear{Pont et al.}{2013}]{Pont+13} Pont F., et al., 2013, MNRAS, 432, 2917                                                       %
\bibitem[\protect\citeauthoryear{Sanchis-Ojeda et al.}{2011}]{Sanchis+11} Sanchis-Ojeda R., et al., 2011, ApJ, 733, 127                                     %
\bibitem[\protect\citeauthoryear{Southworth}{2008}]{Southworth08} Southworth J., 2008, MNRAS, 386, 1644                                                     
\bibitem[\protect\citeauthoryear{Southworth}{2009}]{Southworth09} Southworth J., 2009, MNRAS, 394, 272                                                      
\bibitem[\protect\citeauthoryear{Southworth}{2010}]{Southworth10} Southworth J., 2010, MNRAS, 408, 1689                                                     
\bibitem[\protect\citeauthoryear{Southworth}{2011}]{Southworth11} Southworth J., 2011, MNRAS, 417, 2166                                                     
\bibitem[\protect\citeauthoryear{Southworth et al.}{2009}]{Southworth+09} Southworth J., et al., 2009, MNRAS, 396, 1023                                     
\bibitem[\protect\citeauthoryear{Southworth et al.}{2014}]{Southworth+14} Southworth J., et al., 2014, MNRAS, 444, 776                                      %
\bibitem[\protect\citeauthoryear{Tregloan-Reed et al.}{2013}]{Tregloan+13} Tregloan-Reed J., Southworth J., Tappert C., 2013, MNRAS, 428, 3671              
\bibitem[\protect\citeauthoryear{VandenBerg et al.}{2006}]{VandenBerg+06} VandenBerg D.\ A., Bergbusch P., Dowler P., 2006, ApJS, 162, 375                  
\bibitem[\protect\citeauthoryear{von Essen et al.}{2013}]{Essen+13} von Essen C., Schr\"{o}ter S., Agol E., Schmitt J.\ H.\ M.\ M., 2013, A\&A, 555, A92    

\end{thebibliography}
\end{document}